\documentclass[acmsmall,screen,nonacm]{acmart}

\usepackage{amsmath}
\usepackage{graphicx}
\usepackage{booktabs}
\usepackage{tabularx}
\usepackage{array}
\usepackage{multirow}
\usepackage{makecell}
\usepackage{ragged2e}
\usepackage{pifont}
\hypersetup{
  pdftitle={From AIOps Pipelines to Agentic Service Operations: A Survey of Capability, Authority, and Evidence},
  pdfauthor={Hailiang Zhao et al.},
  pdfsubject={Survey of agentic service operations, AIOps, service systems, failure diagnosis, and root cause analysis},
  pdfkeywords={Agentic service operations, agentic DevOps, AIOps, microservices, Kubernetes, anomaly detection, root cause analysis}
}

\setcopyright{none}
\acmJournal{CSUR}
\newcolumntype{P}[1]{>{\RaggedRight\arraybackslash\hspace{0pt}}p{#1}}
\newcolumntype{Y}{>{\RaggedRight\arraybackslash\hspace{0pt}}X}
\newcommand{\cmark}{\ding{51}}
\newcommand{\xmark}{\ding{55}}

\begin{document}

\title[From AIOps Pipelines to Agentic Service Operations]{From AIOps Pipelines to Agentic Service Operations: A Survey of Capability, Authority, and Evidence}

\author{Hailiang Zhao}
\affiliation{\institution{School of Software Technology, Zhejiang University}\city{Ningbo}\country{China}}
\email{hliangzhao@zju.edu.cn}

\author{Wenzhuo Qian}
\affiliation{\institution{College of Computer Science and Technology, Zhejiang University}\city{Hangzhou}\country{China}}
\email{qwz@zju.edu.cn}

\author{Jianwei Yin}
\affiliation{\institution{College of Computer Science and Technology, Zhejiang University}\city{Hangzhou}\country{China}}
\email{zjuyjw@zju.edu.cn}



\author{Shuiguang Deng}
\authornote{Corresponding author.}
\affiliation{\institution{College of Computer Science and Technology, Zhejiang University}\city{Hangzhou}\country{China}}
\email{dengsg@zju.edu.cn}

\begin{abstract}
Modern service systems combine rapidly changing dependencies, heterogeneous telemetry, and operational actions whose side effects may amplify an incident. Conventional AIOps research has produced effective detectors, localizers, and diagnostic models, yet surveys typically organize them by task, telemetry modality, or learning technique. Such taxonomies describe what a method predicts, but not which decisions it may make or which evidence justifies that authority. We survey the transition from AIOps pipelines to agentic service operations through three coupled lenses: capability, operational authority, and evaluation evidence. We first relate SOA, microservices, Kubernetes, service meshes, and serverless platforms to the incident lifecycle. We then review conventional AIOps methods and audit 22 language-model-enabled systems according to their evaluated behavior rather than their intended autonomy. The audit shows that most systems remain copilots or read-only investigators. Only two exercise write authority; both use constrained action envelopes, and neither carries controlled production-outcome evidence. Building on these findings, we synthesize a four-plane reference architecture, evidence grades, an evaluation ladder, and a controlled adoption path. Together, these artifacts clarify which claims current evaluations support and what evidence is still required for dependable agentic service operations.
\end{abstract}



\maketitle

\section{Introduction}\label{sec:introduction}

Modern online services support commerce, communication, finance, and public infrastructure. To meet growing demand and shorten delivery cycles, software systems have evolved from monolithic applications to service-oriented and cloud-native architectures with independently deployed microservices, containers, Kubernetes platforms, and programmable service meshes~\cite{papazoglou2007soc,dragoni2017microservices,burns2016borg,song2024canalmesh}. This evolution improves scalability and release agility, but it also creates dynamic dependencies, short-lived runtime instances, and long failure-propagation paths. A user-visible timeout or availability drop can originate in application code, a downstream dependency, a deployment, a scheduling decision, a network policy, or retry behavior. Such incidents degrade user experience, disrupt business operations, and can cause substantial losses. Reliable operation consequently depends on timely detection, diagnosis, mitigation, and learning.

The dynamic topology and heterogeneous interactions of service systems make fixed rules and manual scripts insufficient for many investigations. AIOps addresses this problem by applying statistical learning, deep learning, graph models, and causal methods to operational evidence, e.g.,\ logs, metrics, traces, events, topology, and change records~\cite{he2021survey,zhang2026failure,fu2025rcl}. These methods support anomaly detection, alert triage, root-cause localization, diagnosis, and remediation at different stages of incident response. More recent systems can select additional evidence, maintain diagnostic hypotheses, invoke operational interfaces, and propose or execute mitigation under defined controls~\cite{wang2024rcagent,zhang2024flash,chen2025aiopslab,chen2025stratus}. This progression changes the central question. It is no longer sufficient to ask what a model predicts; one must also ask what the system may decide, which tools it can invoke, and what evidence supports that authority.

The service-operations literature spans anomaly detection, root-cause analysis, cloud operations, software engineering, and agent systems. Its distribution across research communities complicates retrieval, comparison, and assessment. Table~\ref{tab:survey-comparison} identifies four gaps in existing surveys. First, reviews often isolate one topic, e.g.,\ log analysis, anomaly detection, root-cause localization, or LLM techniques for AIOps~\cite{he2021survey,kosinska2023observability,zhang2026failure,fu2025rcl,zhang2026llmaiops}, without connecting incident stages in a common operational process. Second, classifications by task, data modality, or model family~\cite{he2021survey,fu2025rcl,zhang2026llmaiops} do not distinguish among recommending a diagnosis, gathering new evidence, and changing production state. These roles entail different decisions and risks. Third, prior reviews rarely examine how service substrates, tool interfaces, permissions, verification, and rollback shape deployment~\cite{zhang2026llmaiops,deng2025asc}. Fourth, existing metrics and benchmarks~\cite{zhang2026failure,fu2025rcl,zhang2026llmaiops} rarely relate an evaluation setting to the authority exercised by a system and to the operational claim that the result can support. Consequently, offline accuracy or sandbox success may be mistaken for evidence of production autonomy.

\begin{table}[htbp]
\centering
\caption{Relationship to representative surveys. A check indicates a primary organizing concern, not merely a mentioned topic. \emph{Authority} records whether the survey organizes work by what a system is permitted to do; \emph{corpus stats} records whether it reports quantitative coverage of its own corpus, e.g.,\ counts by year, venue, or task.}
\label{tab:survey-comparison}
\scriptsize
\setlength{\tabcolsep}{2pt}
\begin{tabularx}{\textwidth}{P{1.66cm}P{2.7cm}*{6}{>{\centering\arraybackslash}p{1.40cm}}}
\toprule
\textbf{Survey} & \textbf{Primary lens} & \textbf{Service arch.} & \textbf{Incident lifecycle} & \textbf{Agent control} & \textbf{Authority} & \textbf{Evidence grade} & \textbf{Corpus stats} \\
\midrule
He et al.~\cite{he2021survey} & Automated log analysis & \xmark & Partial & \xmark & \xmark & \xmark & \cmark \\
Kosi\'{n}ska et al.~\cite{kosinska2023observability} & Cloud-native observability & \cmark & Partial & \xmark & \xmark & \xmark & Partial \\
Zhang et al.~\cite{zhang2026failure} & Microservice failure diagnosis & \cmark & \cmark & \xmark & \xmark & Partial & \cmark \\
Fu et al.~\cite{fu2025rcl} & Root-cause localization & Partial & Partial & \xmark & \xmark & Partial & Partial \\
Zhang et al.~\cite{zhang2026llmaiops} & LLM techniques for AIOps & Partial & \cmark & Partial & Partial & Partial & \cmark \\
Deng et al.~\cite{deng2025asc} & Agentic services computing & \cmark & \xmark & \cmark & Partial & \xmark & \xmark \\
\textbf{This survey} & \textbf{Lifecycle task $\times$ authority} & \cmark & \cmark & \cmark & \cmark & \cmark & Partial \\
\bottomrule
\end{tabularx}
\end{table}

\begin{figure}[tbp]
\centering
\includegraphics[width=\linewidth]{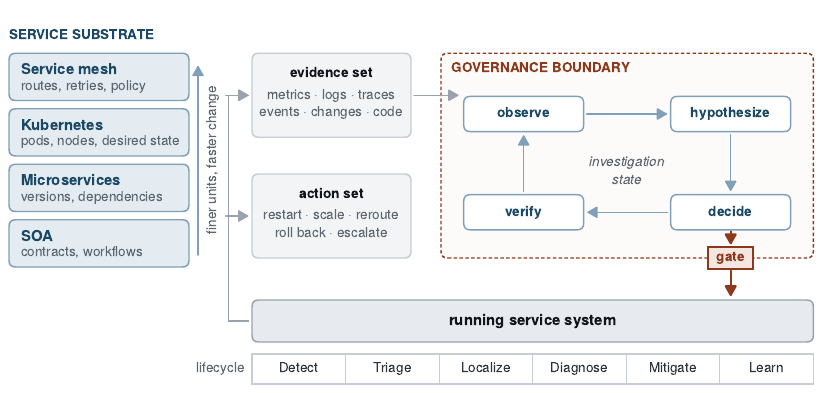}
\caption{Scope of this survey. The service substrate determines which evidence exists and which actions are safe, the incident lifecycle determines the task, and the operational role determines how much of the decision a method assumes. Every state change leaves the governed region through a single gate, which is where operational authority is granted.}
\label{fig:survey-map}
\Description{A ladder of service substrates, from SOA up to service mesh, feeds an evidence set and an action set. The evidence set feeds a four-node loop of observe, hypothesize, decide and verify drawn inside a dashed governance boundary. The write path from the decide node leaves that boundary through a gate and reaches the running service system, which is shown above a ribbon of the six lifecycle stages, from detect through triage, localize, diagnose and mitigate to learn.}
\end{figure}

This survey addresses these gaps by organizing the literature around the operational process and by separating capability from authority and evidence. As illustrated in Figure~\ref{fig:survey-map}, the resulting framework connects the service substrate, incident lifecycle, evidence, and governed action. We use it to compare conventional and language-model-enabled systems, to identify what present evaluations actually establish, and to state the conditions required before greater operational authority is defensible.

The contributions of this survey are as follows:
\begin{itemize}
    \item We establish a common operational problem space that connects service substrates with the incident lifecycle, covering detection, triage, localization, diagnosis, mitigation, and learning.
    \item We introduce a taxonomy of capability, authority, and evidence. It organizes representative conventional systems and audits 22 language-model-enabled systems by their grounding data, tool access, action permissions, evaluation settings, and evaluated authority.
    \item We present a four-plane reference architecture that organizes the evidence, reasoning, tooling, verification, and safety controls required for agentic service operations.
    \item We provide evidence grades, an evaluation ladder, and a research agenda that relate experimental settings to supportable operational claims and guide the progression toward controlled action.
\end{itemize}

The resulting synthesis exposes a clear boundary in the current literature. Fourteen of the 22 audited systems remain at the copilot level, six are read-only investigators, and only two exercise write authority. Both write-capable systems operate within restricted tool or transaction envelopes; one is supported by a user study, the other by interactive testbeds, and neither carries E4 production-outcome evidence. These findings motivate the reference architecture, evaluation ladder, and adoption path developed later in the survey.

The remainder of this article is organized as follows. Section~\ref{sec:method} describes the survey scope, method, authority taxonomy, and evidence grades. Sections~\ref{sec:substrates} through~\ref{sec:evolution} introduce service-system substrates, incidents, the operational problem space, and agentic control loops. Sections~\ref{sec:systems} through~\ref{sec:evaluation} review representative systems, design patterns, and evaluation. The remaining sections discuss open problems, practical adoption, and conclusions.

\section{Scope, Method, and Terminology}\label{sec:method}

A survey is only as trustworthy as the account it gives of its own boundaries. This section states three things a reader needs before the review begins: what we treat as inside the subject and what we deliberately exclude, how the literature was located and filtered, and the two graded vocabularies, for authority and for evidence, that every later claim is expressed in.

\subsection{Scope Boundaries}

This survey examines the runtime operation and evolution of service systems, including monitoring, detection, triage, localization, diagnosis, mitigation, change management, and post-incident learning~\cite{he2021survey,kosinska2023observability,zhang2026failure,fu2025rcl,zhang2026llmaiops}. AIOps refers to data-driven support or automation for these tasks~\cite{he2021survey,dang2019aiops,notaro2021aiops}. An LLM copilot retrieves or generates advice while a human executes and judges it~\cite{zhang2026llmaiops,wang2024agentsurvey}. An operational agent pursues an operational goal through iterative observation and tool use~\cite{yao2023react,qin2024toollearning}. A system is treated as agentic when it exhibits at least two of the following properties: an explicit operational goal, persistent investigation state or hypotheses, iterative selection of observations or tools, and a stopping, escalation, or action decision based on new evidence~\cite{wang2024agentsurvey,yao2023react,qin2024toollearning}.

The scope covers agents that operate a production service system. General coding-agent research is outside this boundary. A coding agent enters the review only when it starts from operational evidence and its proposed change is checked against an operational outcome, as in OrcaRepair~\cite{gao2026orcarepair}. We cite SWE-bench, SWE-agent, and Agentless only where they define relevant evaluation boundaries, not as members of the operational corpus~\cite{jimenez2024swebench,yang2024sweagent,xia2025agentless}. Agentic service operations denotes workflows in which agents coordinate evidence, decisions, and bounded actions under explicit authority~\cite{deng2025asc}. It differs from AgentOps-for-agents, i.e.,\ monitoring and maintaining agent systems themselves. In the terminology of agentic services computing, this survey focuses on operation and governance when the managed environment is a production service system~\cite{deng2025asc}.

\subsection{Evidence-Oriented Literature Selection}

We assembled a curated corpus of work on service-system operations. Searches covered the ACM Digital Library, IEEE Xplore, USENIX proceedings, Proceedings of Machine Learning Research, ICLR/OpenReview, and arXiv. Search strings combined terms for service systems, including microservice and Kubernetes, operational incidents, including AIOps, anomaly detection, failure diagnosis, root cause, and remediation, and agentic methods, including LLM, agent, and multi-agent. Backward and forward citation tracing from relevant surveys and benchmark papers complemented keyword searches~\cite{he2021survey,kosinska2023observability,zhang2026failure,fu2025rcl,zhang2026llmaiops,pham2025rcaeval,cai2026agenticopseval}. The search and metadata verification cutoff was August 2026.

We included peer-reviewed papers and clearly identified preprints that address a service-system operational problem, describe a reproducible method or a sufficiently specified industrial system, and report an evaluation or a design framework relevant to operational agents. We excluded marketing-only claims and papers outside service operations. For every cited paper, we verified its title, authors, year, venue or preprint identifier, and DOI or URL against an official publication, proceedings, institutional page, arXiv, or OpenReview record.

Throughout the review, the five evidence grades in Table~\ref{tab:evidence} distinguish benchmark performance from evidence for operation~\cite{pham2025rcaeval,cai2026agenticopseval,kim2022rigorous}. Grades E1 to E4 label the markers in Figure~\ref{fig:autonomy-matrix} and the final columns of Tables~\ref{tab:agent-systems} through~\ref{tab:agent-systems-c}, allowing a system's lifecycle stage, evaluated authority, and evidence strength to be read together. Grade E0 applies to an unvalidated architecture or proposal cited only for framing and is excluded from the system audit.

\begin{table}[htbp]
\centering
\caption{Evidence grades used throughout this survey. The grade records the strongest evaluation reported for the component under discussion, so it rises only when the experiment becomes more expensive to run, not when the system becomes more capable.}
\label{tab:evidence}
\small
\begin{tabularx}{\textwidth}{P{1.05cm}P{3.6cm}Y}
\toprule
\textbf{Grade} & \textbf{Setting} & \textbf{What it establishes, and what it does not} \\
\midrule
E0 & Concept & An architecture, vision, or unvalidated proposal; establishes feasibility of the design only. \\
E1 & Benchmark & Public or constructed offline data, including fault injection; establishes comparative accuracy, not behavior under organic failures. \\
E2 & Production replay & Historical production incidents or telemetry evaluated offline; adds realistic distributions, but retains hindsight. \\
E3 & Deployment or user study & A user study, a shadow workflow, or a component deployed in production; establishes that people use it, not that outcomes improved. \\
E4 & Controlled outcome & A prospective canary or production comparison measuring recovery and adverse effects; the only grade that establishes an outcome. \\
\bottomrule
\end{tabularx}
\end{table}

Evidence claims apply to the component that was evaluated~\cite{chen2024rcacopilot}. A production data collector does not raise an LLM diagnosis beyond offline evidence when the diagnosis itself was evaluated offline.

\section{Service-System Substrates and Operational Consequences}\label{sec:substrates}

The service substrate determines the objects that an operational method can observe and change. It defines the unit of failure, the relationships recorded between components, the available evidence, and the actions that can be executed safely. A method that ranks services assumes that the service graph is available and current, while a method that restarts an instance assumes that instances are replaceable. These assumptions change across service architectures and limit the transfer of operational methods between them.

SOA organizes capabilities through discoverable interfaces and business workflows~\cite{papazoglou2007soc}. Its operational identity is usually a service, endpoint, or workflow step, and its main evidence comes from contracts, message identifiers, queues, transaction records, and middleware logs. Failures often cross organizational or external-service boundaries, while corrective actions include retrying, rerouting, or compensating a workflow step. Microservices divide these capabilities into smaller independently deployed units~\cite{dragoni2017microservices}. This improves delivery autonomy but increases remote calls, version combinations, ownership boundaries, and partial failures~\cite{taibi2017migrating}. Engineers therefore depend on knowledge of the service graph when locating faults~\cite{zhou2018fault}. Metrics, logs, traces, deployment records, code, and configuration can identify a service, operation, version, or owner, while rollback, isolation, feature control, and scaling provide possible actions. Retries, queues, and asynchronous calls can still make a downstream victim appear more anomalous than the initiating component.

Kubernetes adds workloads, pods, nodes, controllers, labels, and policies to this operational structure~\cite{burns2016borg}. Controllers continuously reconcile desired state, while schedulers and autoscalers change the mapping between logical services and runtime instances. Events, manifests, labels, and resource metrics can expose scheduling, probe, quota, image, node, and policy failures. Corresponding actions include restarting, rescheduling, scaling, or patching a resource, but each action must target the correct level of the service, workload, pod, node, or controller hierarchy. A service mesh further exposes request paths through proxies and a control plane. Distributed tracing supplies request-level dependencies~\cite{sigelman2010dapper}, while cloud-native observability combines traces with metrics, logs, topology, and events~\cite{kosinska2023observability}. The same mesh interfaces can inspect routes, retries, certificates, and policies or change traffic allocation and timeout behavior. The mesh is therefore both a source of evidence and an actuator. Its proxies and control plane can also fail, as shown by efforts to remove sidecars from cloud-scale deployments~\cite{song2024canalmesh}, so read access and write access must remain separate.

Function-as-a-service removes the stable tenant-visible instance and places scheduling and lifecycle control behind a provider boundary~\cite{jonas2019berkeley,shahrad2020serverless}. Operational evidence is largely limited to invocation logs, duration and billing records, trigger metadata, and quotas. Root-cause candidates include the function version, trigger configuration, concurrency limit, downstream quota, and provider infrastructure. Available actions include adjusting concurrency, rebinding a trigger, or rolling back a function version, while restarting a specific instance is generally unavailable. The decisive evidence may also remain inside the provider control plane. Across these substrates, an operational agent therefore requires a versioned resource model that connects business capabilities with services, versions, workloads, instances, nodes, routes, owners, and service objectives. Its hypotheses, tool calls, approvals, and outcomes become additional operational records, and its actions add another possible failure path. The next section formalizes the observations, dependency structure, lifecycle outputs, and permitted actions defined by these substrate-level constraints.

\section{The Operational Problem Space}\label{sec:problem}

Methods from anomaly detection, graph learning, causal inference, and language modeling are usually evaluated within their own research communities. This section places them in a common operational frame. It first distinguishes the objects involved in an incident and explains why distributed diagnosis requires a sequential investigation. It then formalizes the available observations and the six stages of the incident lifecycle, before describing how an operational agent selects further observations and actions.

\subsection{Incidents as Sequential Operational Problems}

A fault is a condition capable of producing incorrect behavior, an error is the resulting incorrect internal state, and a failure is a deviation of delivered service from its specification~\cite{avizienis2004taxonomy}. A symptom is an observable manifestation of an error, whereas impact describes its consequence for users or business objectives. An incident is an interval in which service behavior violates or credibly threatens an operational objective. Service-level objectives and error budgets provide explicit criteria for making this judgment~\cite{beyer2016site}. Classic studies of fault-tolerant and internet services identify software faults and operator actions as prominent outage sources~\cite{gray1986stop,oppenheimer2003why}.

This survey defines a root cause as the earliest condition in the investigated causal chain on which an operator could have acted. It need not be the first event in time or the component with the largest anomaly score. Several conditions may jointly produce an incident, so diagnosis may need to identify multiple actionable causes~\cite{cook1998complex}. Consider a payment-service release that retains objects in an unbounded cache. Memory growth causes garbage-collection stalls, slow payment calls trigger checkout retries, and the additional traffic increases latency until an SLO is violated. The alert is a symptom, abandoned orders represent the impact, and the release configuration is an actionable cause. Restarting a pod may temporarily reduce latency without removing the fault. Rolling back the release removes the causal condition, but recovery is complete only after the SLO and downstream queues have returned to an acceptable state.

Distributed service incidents require investigation over time rather than classification from a static snapshot. Fan-out, queues, retries, circuit breakers, and autoscaling can move symptoms away from their causes~\cite{dragoni2017microservices}. Service, version, workload, pod, node, route, and owner are separate operational entities whose relationships change during scaling and deployment~\cite{burns2016borg,wang2026topoevo,huang2026star}. Trace sampling, metric aggregation, missing logs, and clock differences further leave evidence incomplete or misaligned, motivating methods that reason over missing metrics and call-graph blind spots~\cite{zhang2025reconrca}. Operational actions also change the system state and the evidence available to subsequent decisions.

An operational method that interacts with the service should therefore produce an investigation trajectory rather than a single answer. The trajectory records observations, hypotheses, tool calls, decisions, and outcome checks. It should distinguish observed facts tied to a source and time interval, inferences with stated uncertainty, proposed probes that separate competing explanations, and action contracts that define the permitted change and its success conditions. A system can advance an investigation by collecting discriminating evidence or rejecting an unsafe action even when it does not identify a root cause or hold permission to change production state. The following subsection formalizes the observations, lifecycle outputs, and actions that constitute this trajectory.

\subsection{Operational Evidence and Lifecycle Tasks}

At time $t$, an operational method observes a history $O_{0:t}$ rather than the service state directly. The evidence includes metrics, logs, traces, platform and application events, code and configuration, deployment records, and operational knowledge, e.g.,\ service catalogs, runbooks, and incident history. It also includes the dynamic dependency graph $G_t$, which relates services, workloads, instances, nodes, routes, and owners. These sources differ in latency, sampling, semantic stability, and access control. Cloud-native observability therefore depends on correlating heterogeneous evidence~\cite{kosinska2023observability}, while distributed tracing provides the request-level relationships needed to reconstruct propagation paths~\cite{sigelman2010dapper}.

\begin{figure}[tbp]
    \centering
    \includegraphics[width=\linewidth]{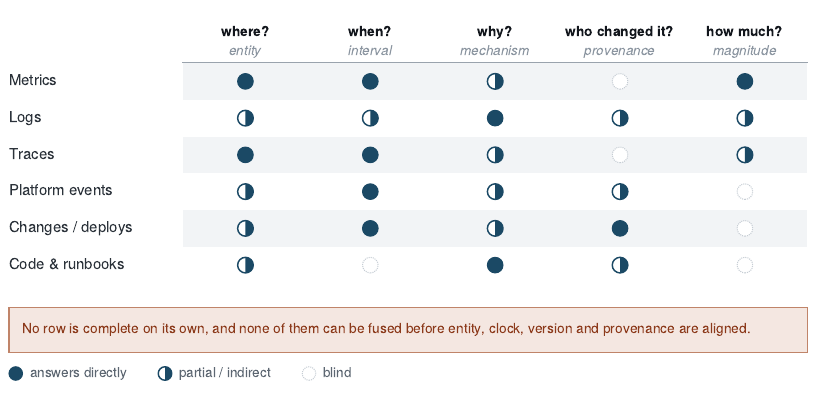}
    \caption{Operational evidence is complementary. Each modality answers some questions directly or indirectly and leaves others unresolved. Combining them requires alignment of entity, time, version, and provenance.}
    \label{fig:telemetry}
    \Description{A matrix whose rows represent metrics, logs, traces, platform events, deployments and changes, and code and operational knowledge. Its columns represent the questions where, when, why, who changed it, and how much. Filled, half-filled, and dotted circles indicate direct, partial, and absent coverage.}
\end{figure}

Figure~\ref{fig:telemetry} summarizes the coverage of each evidence source. Completeness concerns evidence lost through sampling, missing instrumentation, or omitted logs. Fidelity concerns clock differences, aggregation, noisy labels, and incorrect topology. Timeliness concerns evidence that arrives after an operational decision must be made. Reasoning cannot recover information that was never observed, so a system must either request another measurement or report that the available evidence does not distinguish its remaining hypotheses.

Operational evidence supports six stages with different objectives and outputs. Detection is commonly studied through log and metric analysis~\cite{he2021survey,chandola2009anomaly}, triage through incident-management systems~\cite{chen2020incidental}, localization through microservice root-cause methods~\cite{fu2025rcl}, and diagnosis across the broader failure-management process~\cite{zhang2026failure,notaro2021aiops}. Table~\ref{tab:lifecycle} defines these stages and their evaluation criteria. No single accuracy measure represents the entire lifecycle because useful performance depends on the output, decision time, and cost of an incorrect action~\cite{salfner2010failure}.

\begin{table}[htbp]
\centering
\caption{Outputs and common metrics across the service incident lifecycle.}
\label{tab:lifecycle}
\small
\begin{tabularx}{\textwidth}{P{1.45cm}P{7.0cm}Y}
\toprule
\textbf{Stage} & \textbf{Primary output} & \textbf{Common metrics} \\
\midrule
Detect & Determines whether service behavior is abnormal and returns the affected interval and severity. & Precision, recall, F1, detection delay, and false-alert rate. \\
Triage & Groups related alerts into incidents, assigns a known failure class and severity, and routes each incident to an owner. & Grouping purity, classification accuracy, routing accuracy, and time to assign. \\
Localize & Ranks the services, instances, changes, metrics, or code regions most likely to contain the cause. & Top-$k$ accuracy, MRR, and rank stability. \\
Diagnose & Explains the failure mechanism as a causal account with supporting evidence and uncertainty. & Cause accuracy, evidence precision, contradiction rate, calibration, and human usefulness. \\
Mitigate & Selects or proposes a plan, playbook, bounded action, or escalation that restores the service objective. & Recovery success, mitigation time, SLO recovery time, side-effect rate, and rollback success. \\
Learn & Converts a verified incident outcome into a runbook, regression test, monitor, or postmortem update. & Recurrence rate, regression coverage, knowledge freshness, and toil reduction. \\
\bottomrule
\end{tabularx}
\end{table}

The stages are connected but not interchangeable. Triage assigns an incident to a known class and owner, localization ranks where to investigate, and diagnosis explains the mechanism using supporting evidence. A system may perform well at one stage and still provide an unsuitable input to the next. Mitigation must therefore be evaluated by recovery and adverse effects, while learning must be evaluated by whether a later incident becomes easier to resolve.

\subsection{Agentic Formulation}

An operational agent extends this lifecycle by selecting its next observation or action from the current evidence. Let $B_t$ denote its state over candidate causes, $K_t$ its operational knowledge, $\mathcal{T}$ its typed tool catalog, and $\mathcal{P}$ its policy and permission set. Its decision can be written as
\begin{equation}
d_t \sim \pi(B_t,O_{0:t},G_t,K_t,\mathcal{T},\mathcal{P}),
\qquad
d_t=a_t \Longrightarrow a_t\in A_t^{\mathrm{allowed}},
\label{eq:agent-decision}
\end{equation}
where $d_t$ may be a query, test, action, escalation, or stop decision, and $A_t^{\mathrm{allowed}}$ is the action set permitted without a new human decision. After a tool returns new evidence, the agent updates its candidate causes and selects the next step. Its objective is to restore the operational target while controlling investigation time, tool cost, and action risk. Additional observations can reduce model uncertainty, but they cannot relax policy or permission constraints. The next section uses this distinction to explain the transition from fixed AIOps pipelines to agentic control loops.

\section{From AIOps Models to Agentic Control Loops}\label{sec:evolution}

This section explains how operational intelligence progresses from models that score prepared inputs to systems that select observations and actions during an incident. It first distinguishes three generations of operational intelligence, then introduces the mechanisms required for conditional control through a four-plane reference architecture. It finally defines operational authority by the strongest behavior a system performs without a new human decision. Together, these concepts support the task-centered audit in Section~\ref{sec:systems} and the evidence-oriented evaluation in Section~\ref{sec:evaluation}.

\subsection{Three Generations of Operational Intelligence}

The first generation converts telemetry into anomaly scores, predictions, or labels. Classical anomaly detection provides the statistical basis for identifying deviations~\cite{chandola2009anomaly}, while failure prediction shows that temporal models can anticipate some failures before they occur~\cite{salfner2010failure}. Later work learned patterns from logs, metrics, and traces. These methods are efficient and widely deployable, but they usually return a signal that still requires an operator to connect symptoms, identify causes, and choose a response~\cite{lou2013software,dang2019aiops,notaro2021aiops}.

The second generation incorporates service structure. Graph-based methods propagate evidence over dependencies, multimodal methods combine complementary telemetry, and causal methods seek to separate initiating faults from downstream symptoms. These approaches improve localization and diagnosis, but most remain fixed pipelines in which a predefined input representation is passed to a predefined model.

\begin{figure}[tbp]
\centering
\includegraphics[width=\linewidth]{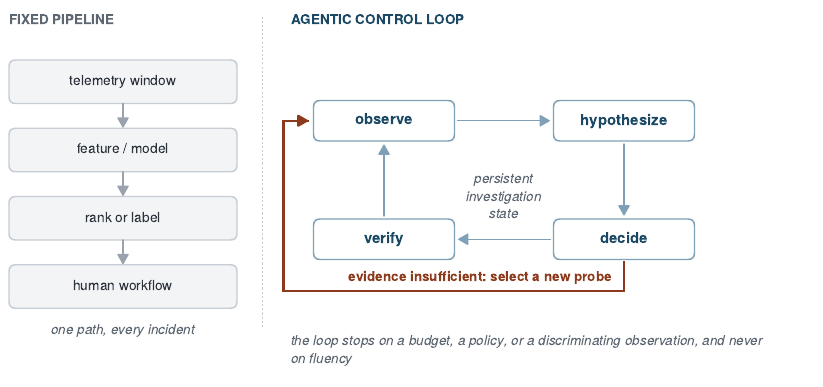}
\caption{A fixed pipeline follows the same path for every incident. An agentic loop maintains investigation state and selects a new probe when current evidence is insufficient. This conditional return from decision to observation distinguishes an operational agent from a fixed prediction pipeline.}
\label{fig:pipeline-vs-agent}
\Description{On the left, a four-stage chain runs from a telemetry window through a model to a rank or label and then to a human workflow. On the right, a four-node cycle of observe, hypothesize, decide and verify surrounds persistent investigation state, with an additional highlighted edge returning from decide to observe when evidence is insufficient.}
\end{figure}

The third generation makes the investigation path conditional on intermediate evidence. As Figure~\ref{fig:pipeline-vs-agent} shows, an operational agent can query traces after detecting latency, inspect a deployment change when several services remain plausible causes, or stop and escalate when available evidence is insufficient. The defining property is therefore not natural-language generation, but the ability to select the next observation or bounded action from the current investigation state. This development extends the monitor, analyze, plan, execute, and knowledge structure of autonomic computing~\cite{kephart2003autonomic,huebscher2008autonomic}. Language models can support semantic interpretation and flexible tool selection, while telemetry models, graph algorithms, policy engines, and validators remain necessary parts of the operational loop.

\subsection{Agent Building Blocks in Plain Language}

Conditional control requires several distinct mechanisms. General agent research commonly separates memory, planning, and action~\cite{wang2024agentsurvey,xi2025rise}; operational systems require a more explicit version of this separation because their evidence and actions concern a changing production system. Grounding binds a claim to an entity, version, time interval, baseline, and source. A world model records the current topology, ownership, desired state, and recent changes. Investigation state preserves competing hypotheses, their support and contradictions, so that early plausible explanations do not determine all later queries. Memory stores incident records and reusable procedures, but retrieved knowledge must retain its scope and freshness because a past procedure can become unsafe after the system changes.

A planner selects the next probe, action, escalation, or stopping point. Stepwise reasoning, multiple candidate paths, and search over partial solutions provide useful planning mechanisms~\cite{wei2022cot,wang2023selfconsistency,yao2023tot}, but operational planning must remain tied to observable evidence. Tools therefore expose typed read, test, and write operations rather than unrestricted commands~\cite{schick2023toolformer,qin2024toollearning}. A validator checks evidence provenance, parameters, policy, preconditions, and outcomes. Model self-reflection can help identify inconsistencies~\cite{shinn2023reflexion}, but service-state validation remains necessary because a coherent explanation or executable command does not establish a correct diagnosis or safe mitigation. These mechanisms map directly to the four planes in Figure~\ref{fig:reference-architecture}.

\subsection{A Reference Architecture for Operational Agents}

\begin{figure}[tbp]
\centering
\includegraphics[width=\linewidth]{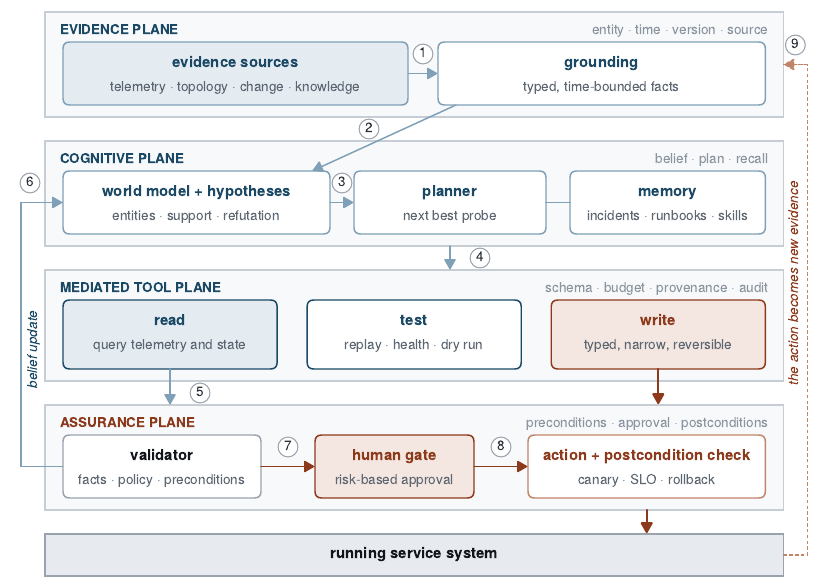}
\caption{Reference architecture for an operational agent. The evidence plane grounds operational records, the cognitive plane maintains investigation state and selects the next step, the tool plane mediates typed read, test, and write operations, and the assurance plane validates evidence and actions. The read path forms the L2 investigation loop; every write passes through the assurance gate and returns as new evidence.}
\label{fig:reference-architecture}
\Description{Four stacked horizontal bands, labelled evidence plane, cognitive plane, mediated tool plane and assurance plane, sit above a band representing the running service system. The evidence plane holds evidence sources and a grounding stage; the cognitive plane holds a world model with hypotheses, a planner and a memory; the tool plane holds separate read, test and write lanes governed by schema, budget, provenance and audit; the assurance plane holds a validator, a human gate and an action stage with postcondition checks. Numbered arrows pass between the bands. A left-hand lane returns a belief update from the validator to the world model, and a right-hand dashed lane returns the executed change from the running system to the evidence plane.}
\end{figure}

Figure~\ref{fig:reference-architecture} presents a reference architecture derived from tool-augmented root-cause analysis, SOP-guided diagnosis, runbook assistance, and interactive evaluation environments~\cite{wang2024rcagent,pei2025flow,an2024nissist,chen2025aiopslab}. It organizes an operational agent into four planes.

The \emph{evidence plane} connects telemetry, topology, code, configuration, changes, and operational knowledge, then grounds raw records as stable and time-bounded facts. The \emph{cognitive plane} maintains the world model, candidate hypotheses, memory, and a planner for selecting the next probe. The \emph{tool plane} provides typed read, test, and write interfaces through a broker that can enforce schemas, budgets, provenance, and permissions. The \emph{assurance plane} checks contradictions, action preconditions, approval requirements, postconditions, and rollback conditions. Human responders participate through this plane by approving, correcting, or escalating decisions.

The separation identifies where an operational system may fail. Weak grounding produces claims from mismatched entities or time windows. Weak hypothesis management commits the investigation to an early explanation. Unmediated tools bypass policy and audit controls. Weak assurance converts an incorrect diagnosis into an unsafe change. The architecture therefore provides a common vocabulary for the system audit in Section~\ref{sec:systems} and the design patterns in Section~\ref{sec:patterns}.

\subsection{Autonomy Is an Authority Property}

The four planes describe how an operational system is organized, but not how much authority it holds. Table~\ref{tab:autonomy} therefore defines five levels according to the strongest behavior a system performs without a new human decision. This definition follows human-factors research that separates the stage of information processing from the authority granted to automation~\cite{sheridan1978undersea,parasuraman2000levels}. It also reflects established risks of misuse, disuse, loss of operator skill, and poorly calibrated trust~\cite{parasuraman1997use,bainbridge1983ironies,lee2004trust,hoffman2013trust}.

A system that ranks suspect services remains an offline analytic method even when it uses multiple models. A system that retrieves evidence and recommends a diagnosis is a copilot when a human chooses and executes every action. A read-only investigator becomes more capable by selecting its own diagnostic queries, but it does not gain write authority. Supervised and closed-loop operation require explicit action boundaries, verification, rollback, and human responsibility similar to the conditions used to define driving automation~\cite{sae2021j3016}. The authority level thus depends on actual operational behavior, not on model size, natural-language fluency, or the number of collaborating agents.

\begin{table}[htbp]
\centering
\caption{Operational authority levels. Advancement requires stronger assurance, not only a stronger model.}
\label{tab:autonomy}
\small
\begin{tabularx}{\textwidth}{P{2.35cm}P{3.55cm}P{4.3cm}Y}
\toprule
\textbf{Level and role} & \textbf{Permitted behavior} & \textbf{Human responsibility and minimum assurance} & \textbf{Representative form} \\
\midrule
L0: Offline analytic & Score, classify, or rank supplied data. & Human interprets and acts; dataset validity and calibration are required. & Detector or RCA model. \\
L1: Copilot & Retrieve, summarize, explain, and recommend. & Human selects evidence and executes; citations, provenance, and uncertainty are required. & Incident summary or RCA assistant. \\
L2: Read-only investigator & Iteratively query telemetry and knowledge. & Human approves conclusions and all writes; typed reads, budgets, audit, and a stop rule are required. & Tool-augmented RCA agent. \\
L3: Supervised operator & Prepare or execute an explicitly approved bounded action. & Human authorizes the transition; preconditions, preview, approval, and rollback are required. & Supervised runbook executor. \\
L4: Bounded closed loop & Detect, investigate, and execute pre-approved reversible actions. & Human defines policy and reviews exceptions; action envelopes, canaries, postconditions, and a kill switch are required. & Self-healing controller with agentic diagnosis. \\
\bottomrule
\end{tabularx}
\end{table}

\begin{figure}[tbp]
\centering
\includegraphics[width=\linewidth]{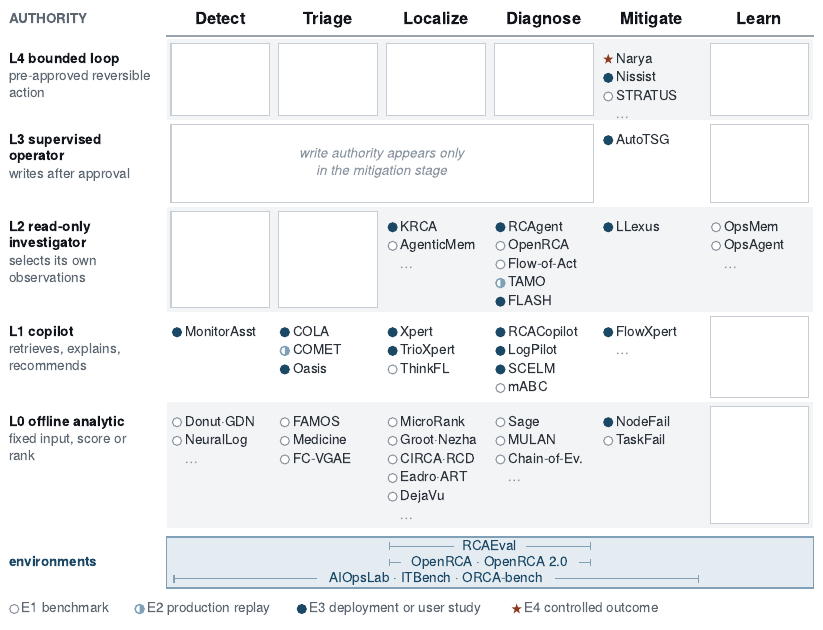}
\caption{Capability, evaluated authority, and evidence across the reviewed literature. Columns represent lifecycle stages and rows represent the strongest authority exercised during evaluation. Markers show the strongest evidence grade. Most systems occupy analytic, copilot, or read-only investigation roles; evaluated write authority is concentrated in mitigation, and learning remains confined to read-only use of prior incident records.}
\label{fig:autonomy-matrix}
\Description{A matrix with detection, triage, localization, diagnosis, mitigation and learning as columns and the five authority levels L0 to L4 as rows. Cells are populated with named systems, each carrying a marker for evidence grade E1 to E4. The lower-left region is dense, the L3 and L4 rows are empty except in the mitigation column, and the learning column carries entries at the L2 read-only level alone. A band beneath the matrix shows which columns the evaluation environments cover.}
\end{figure}

Figure~\ref{fig:autonomy-matrix} places the reviewed literature along two independent dimensions. The horizontal axis records lifecycle coverage from detection to learning. The vertical axis records the authority exercised during evaluation. A method can provide sophisticated diagnosis while holding only read authority, whereas a bounded controller can execute one pre-approved recovery action without general diagnostic ability. The matrix therefore records evaluated behavior rather than the authority an architecture might permit in principle. It also shows that evaluated write authority is concentrated in mitigation and that learning systems remain limited to read-only use of incident history. The next section uses these two axes to review representative systems and compare their capabilities with the evidence available for their operational claims.

\section{Representative Systems in Service Operations}\label{sec:systems}

The previous section defined the operational objects, lifecycle outputs, authority levels, and evidence grades used in this survey. This section applies that framework to representative systems. We first review analytical outputs, then examine copilot and agentic investigation, governed change, and operational learning. Figure~\ref{fig:autonomy-matrix} and Table~\ref{tab:traditional-capabilities} summarize the analytical landscape, while Tables~\ref{tab:agent-systems} through~\ref{tab:agent-systems-c} audit language-model-enabled systems by grounding data, interaction, authority, and evidence.

\subsection{Operational Outputs in Incident Handling}

Operational systems are first distinguished by their output. Detection identifies abnormal behavior and its interval, triage groups and labels related signals, localization ranks suspect entities, and diagnosis explains the failure with supporting evidence. Each output narrows the next decision but does not replace it.

Log detectors progressed from template extraction and sequence modeling to robust and parser-free representations. Spell, Drain, and DeepLog established the basic pipeline~\cite{du2016spell,he2017drain,du2017deeplog}; LogRobust and NeuralLog address drift or parsing cost~\cite{zhang2019logrobust,le2021neurallog}. LogHub~2 shows that parser accuracy and method ranking change on realistic data~\cite{jiang2024loghub2}, so deployment context remains part of the detection problem.

Metric methods cover seasonal and multivariate behavior with autoencoder, recurrent, transformer, and graph models~\cite{xu2018donut,su2019omni,tuli2022tranad,deng2021gdn}. Trace methods preserve path structure through graph-based representations; TraceBench shows that their ranking is sensitive to data and preprocessing~\cite{sun2025tracebench}.

Multimodal systems combine complementary evidence through shared graphs or cross-modal representations, as in Eadro, ART, and FAMOS~\cite{lee2023eadro,sun2024art,duan2025famos}. Their main operational limits are modality imbalance and missing data, addressed by adaptive weighting and uncertainty-aware detection~\cite{tao2024everymodality,sun2026uncertainty}. FAMOS illustrates an important boundary: its output is a failure class, so it belongs to triage rather than localization.

For language models, direct detection on logs or time series has mixed gains and high cost. More practical designs use the model to write rules or scope evidence for cheaper detectors~\cite{gu2025argos,jiang2025logpilot}. Thus, specialized models remain preferable for high-frequency scoring, while language models add value in interpreting alerts, selecting intervals, linking changes and runbooks, and choosing the next analysis.

Incident formation groups alert streams into actionable incidents. Because one fault can trigger alerts across dependent components~\cite{chen2020incidental}, aggregation, failure typing, assignment, and incident summarization precede localization. Errors at this stage hide discriminating evidence and delay ownership.

COLA combines correlation with LLM tie-breaking for alert aggregation~\cite{kuang2024cola}. FAMOS, Medicine, and graph autoencoders classify known failure types from multimodal or graph data~\cite{duan2025famos,tao2024everymodality,sun2026failurevgae}; their output is a label $\tau$, not a causal account. COMET predicts an owner with an explanation~\cite{wang2024comet}, while Oasis summarizes the incident state and Xpert recommends the next query~\cite{jin2023oasis,jiang2024xpert}. MonitorAssistant turns monitor configuration into a maintainable artifact~\cite{yu2024monitorassistant}, and LLexus combines several steps while retaining a human gate for service changes~\cite{lascasas2024llexus}. These systems show that triage is a distinct operational output and a prerequisite for higher-authority action.

The audited triage systems are mostly supported by deployment or incident-corpus evidence, but remain at read-only authority. They provide the incident representation on which later investigation and any governed change depend.

Localization ranks candidate services, instances, metrics, or code regions. Trace and graph methods exploit execution structure and anomaly propagation~\cite{yu2021microrank,gan2023sleuth,yu2023nezha}. Their central trade-off is search cost versus graph coverage.

The reported granularity has moved from service ranking toward metric and code localization. Finer outputs can be more actionable, but only when the underlying graph and evidence cover the fault.

Diagnosis adds a causal account to localization by distinguishing interventions from propagated symptoms. The literature spans temporal, Bayesian, multimodal, event, and latent-variable models~\cite{gan2021sage,li2022circa,ikram2022rcd,zheng2024mulan,yao2024coe,xie2024latentscope}. Its assumptions about noise, caller ambiguity, component independence, propagation lag, and topology stability must remain explicit.

A recurring failure can be recognized more cheaply than it can be re-derived. DejaVu uses incident history to return known failure units~\cite{li2022dejavu}; history is therefore an input to localization, not evidence of operational learning.

Groot shows why production systems often prefer auditable event graphs to opaque models~\cite{wang2021groot}. Comparative evaluation likewise finds no universal causal-RCA winner and weak transfer from synthetic to real systems~\cite{pham2024causal}; causal graphs should therefore serve as priors and explanation structures whose assumptions remain inspectable.

These systems produce complementary outputs: detection and triage establish the incident, localization narrows candidates, and diagnosis supplies an evidence-backed causal account. Table~\ref{tab:traditional-capabilities} summarizes the capabilities that an operational agent must coordinate.

\begin{table}[htbp]
\centering
\caption{Representative pre-agentic AIOps capabilities and what an agent still has to supply.}
\label{tab:traditional-capabilities}
\small
\begin{tabularx}{\textwidth}{P{4.0cm}P{4.6cm}Y}
\toprule
\textbf{Capability family and representative work} & \textbf{Contribution and operational assumption} & \textbf{Remaining orchestration gap} \\
\midrule
Log sequences: Drain~\cite{he2017drain}; DeepLog~\cite{du2017deeplog}. & Extract stable events and deviations from normal order, assuming logs are parsable and adequately emitted. & Select incident scope and connect log anomalies to other evidence. \\
Metric dependencies: OmniAnomaly~\cite{su2019omni}; GDN~\cite{deng2021gdn}. & Produce multivariate temporal or graph anomaly scores, assuming training data represents normal dependencies. & Interpret business impact and a topology that changes after training. \\
Multimodal state: Eadro~\cite{lee2023eadro}; ART~\cite{sun2024art}; FAMOS~\cite{duan2025famos}. & Learn complementary system representations, assuming modalities can be aligned and training faults are representative. & Decide which evidence is sufficient and which observation to request next. \\
Trace/graph localization: MicroRank~\cite{yu2021microrank}; Sleuth~\cite{gan2023sleuth}; Nezha~\cite{yu2023nezha}. & Rank components using dependency structure and interpretable patterns, assuming traces reflect propagation. & Reconcile conflicting ranks and changes outside the observed graph. \\
Causal RCA: CIRCA~\cite{li2022circa}; RCD~\cite{ikram2022rcd}; Chain-of-Event~\cite{yao2024coe}; LatentScope~\cite{xie2024latentscope}. & Offer intervention-oriented explanations, assuming adequate graph quality, event history, or causal sufficiency. & Test causal assumptions and actively seek counterevidence. \\
Adaptive mitigation: Narya~\cite{levy2020narya}. & Combines prediction with preventive VM mitigation where a bounded action set and measurable outcome exist. & Generalize beyond a narrow action envelope while preserving outcome safety. \\
\bottomrule
\end{tabularx}
\end{table}

\subsection{Agent Participation in Investigation}

The systems in this part participate in the outputs above rather than adding new lifecycle stages. Copilots retrieve, explain, and recommend. Active investigators choose observations, invoke tools, maintain hypotheses, and stop or escalate when evidence is insufficient. Interaction and authority distinguish the two roles.

Copilot systems synthesize supplied evidence while humans retain the decision. Ahmed et al.\ evaluate incident-title- and summary-based recommendations on more than 40{,}000 Microsoft incidents~\cite{ahmed2023recommending}. RCACopilot adds a predefined diagnostic collector before LLM classification and explanation~\cite{chen2024rcacopilot}, and GPT-4 in-context learning extends the same setting to more than 100{,}000 incidents~\cite{zhang2024iclrca}. These systems improve access to prior knowledge but remain L1 advice.

COCA retrieves code and execution context when runtime telemetry is absent~\cite{li2025coca}; SCELM combines multimodal change evidence with retrieval for change assessment and RCA~\cite{sun2025scelm}. The common value is traceable synthesis of fragmented evidence, not autonomous action.

Active investigators select observations and use tools, moving from L1 advice to L2 read-only investigation. RCAgent integrates tool selection, context management, and domain knowledge in an Alibaba Cloud workflow~\cite{wang2024rcagent}. OpenRCA evaluates tool-assisted agents on 335 failures and more than 68~GB of telemetry, with a best reported solve rate of 11.34\%~\cite{xu2025openrca}. The result places long-context retrieval and telemetry coverage, rather than abstract reasoning, at the center of the problem.

A common design constrains the search with procedures. Flow-of-Action embeds SOPs in a ReAct loop and reports 64.01\% localization accuracy versus 35.50\% for unconstrained ReAct~\cite{pei2025flow}. FlowXpert evolves workflows from operational documents~\cite{shi2025flowxpert}. Procedures supply the search structure; the model handles the remaining judgments.

Using multiple models or roles does not by itself establish L2 agency. TrioXpert specializes processing by modality, mABC follows a predefined collaboration and voting workflow, and ThinkFL changes the training objective~\cite{sun2025trioxpert,zhang2024mabc,zhang2026thinkfl}. All three operate on supplied evidence and therefore remain L1 in the audit. By contrast, TAMO invokes specialized analysis tools, while KRCA and OpsAgent select graph drilldowns or coordinate evolving investigations~\cite{zhang2025tamo,jiang2026krca,luo2025opsagent}.

A third line supplies structure for causal reasoning. Cloud Atlas induces a graph from documentation and code, TopoEvo tracks topology drift, Praxis joins service and program graphs, CausalDX handles cascading incidents, and SpecRCA verifies hypotheses in parallel~\cite{xie2024cloudatlas,wang2026topoevo,cui2026praxis,chang2026causaldx,zhang2026specrca}. FLASH and X-lifecycle reuse recurring workflows and lifecycle records~\cite{zhang2024flash,goel2024xlifecycle}; ticket retrieval and prompt optimization improve evidence assembly~\cite{isaza2024ragincident,goel2025earco}. Network-operation studies reach the same conclusion that evidence assembly is the bottleneck~\cite{hamadanian2023holistic,wang2024netassistant}, while AIDA and NetAgentMemory add production-oriented verification and shared memory~\cite{xu2026aida,ren2026netagentmemory}.

In database operations, D-Bot combines manual-derived knowledge, tools, and tree search~\cite{zhou2024dbot}. Its better-instrumented environment provides a useful comparison between reasoning limits and observability limits.

Production comparisons attribute much of an agent's advantage to evidence gathering rather than reasoning over identical inputs~\cite{roy2024llmagents}; operational studies likewise identify the query, action, and audit interface as more consequential than the prompt~\cite{shetty2024autonomousclouds}. Open-source systems show demand but not efficacy without comparable evaluation~\cite{k8sgpt2026,holmesgpt2024}. Multi-agent designs are most defensible when they separate evidence processing from deliberation or assign an explicit verification role.

Agentic investigation coordinates evidence and decisions, but does not itself grant permission to change a service. Systems that cross this boundary are discussed next.

\subsection{Governed Operational Change}

A mitigation system connects a diagnosis to a bounded change. Runbooks specify the action, governance constrains target, parameters, permissions, preconditions, approval, and rollback, and execution produces an outcome against which recovery and side effects can be checked.

AutoTSG translates troubleshooting guides into executable workflows, while Nissist retrieves and supports semi-automated execution from guides and mitigation history~\cite{shetty2022autotsg,an2024nissist}. Both show that executable control flow, parameters, preconditions, and tool semantics matter as much as retrieval.

Pre-agentic work shows how bounded actions make outcomes measurable. Configuration studies motivate versioned and validated control surfaces~\cite{xu2015knobs,tang2015holistic,mehta2020rex}; failure prediction couples forecasts with preventive migration at host or workload level~\cite{xu2018disk,lin2018nodefailure,gao2020taskfailure}; canary analysis uses live-traffic evidence to abort unsafe releases~\cite{khande2025canary}.

LLM-generated playbooks show script-level correctness but not production safety~\cite{sarda2024remediation}. OrcaRepair ties code changes to observable fault signatures, and Honeypoint selects responses from constrained placement actions~\cite{gao2026orcarepair,zhang2026honeypoint}. Narya remains the strongest comparison, combining prediction with bounded preventive mitigation and a fifteen-month production outcome~\cite{levy2020narya}. The evidence is strong because the action space and safety conditions are narrow.

\begin{table}[htbp]
\centering
\caption{Capability and evidence audit of language-model-enabled operational systems, part~1 of~3. Inclusion requires an evaluated service-system task from which both lifecycle stage and authority can be read. ``Evaluated authority'' records the strongest tested behavior, and systems tested at adjacent levels are counted at the higher one. Parts~2 and~3 continue in Tables~\ref{tab:agent-systems-b} and~\ref{tab:agent-systems-c}.}
\label{tab:agent-systems}
\scriptsize
\begin{tabularx}{\textwidth}{P{1.45cm}P{2.45cm}P{3.30cm}P{2.30cm}Y}
\toprule
\textbf{Stage} & \textbf{System and venue} & \textbf{Capability and grounding} & \textbf{Evaluated authority} & \textbf{Strongest evidence and boundary} \\
\midrule
Diagnose & Ahmed et al.~\cite{ahmed2023recommending}; ICSE 2023 & Root cause and mitigation from incident title and summary alone. & Fine-tuned generation; L1 advice. & E2: 40{,}000+ production incidents; no telemetry access or investigation. \\
Triage & COLA~\cite{kuang2024cola}; ICSE-SEIP 2024 & Alert aggregation from alert attributes and topology. & Correlation with LLM tie-breaking; L1 support. & E3: reported production alert volumes; aggregation only. \\
Triage & COMET~\cite{wang2024comet}; ISSRE 2024 & Incident triage with an interpretable assignment rationale. & LLM classification plus explanation; L1 advice. & E2/E3: production incident corpus; routing only. \\
Localize & Xpert~\cite{jiang2024xpert}; ICSE 2024 & Recommendation of the next diagnostic query. & Query synthesis from incident context; L1 advice. & E3: reported deployment; recommends rather than executes. \\
Detect & MonitorAssistant~\cite{yu2024monitorassistant}; FSE Comp. 2024 & Monitor configuration and alert explanation. & LLM-assisted configuration; L1 support. & E3: reported deployment; acts on monitors, not on the service. \\
Diagnose & RCACopilot~\cite{chen2024rcacopilot}; EuroSys 2024 & Root-cause category and explanation from incident handlers, diagnostic data, and history. & Predefined collection then LLM inference; L1 advice. & E2/E3: one year of incidents and a deployed collector, but no action authority. \\
Diagnose & GPT-4 ICL RCA~\cite{zhang2024iclrca}; FSE Comp. 2024 & Root causing from more than 100{,}000 production incident examples. & In-context generation; L1 advice. & E2: large production corpus, but no live investigation. \\
Mitigate & LLM remediation~\cite{sarda2024remediation}; FSE Comp. 2024 & Ansible playbook generation from tasks and examples. & In-context or fine-tuned generation; evaluated at L1, artifact targets L3. & E1: functional script tests; generation is not production-safe execution. \\
Diagnose & COCA~\cite{li2025coca}; ICSE 2025 & RCA from issue reports, retrieved code, execution paths, and history. & Retrieval and generation; L1 advice. & E1: five distributed systems; no live telemetry or action. \\
Diagnose & LogPilot~\cite{jiang2025logpilot}; ASE 2025 & Alert-oriented diagnosis from alert logic and request log chains. & Intent scoping, clustering, LLM diagnosis; L1. & E2/E3: production alerts and reported deployment; log-centered scope. \\
\bottomrule
\end{tabularx}
\end{table}

Tables~\ref{tab:agent-systems-b} and~\ref{tab:agent-systems-c} complete the audit. They separate fixed multi-model pipelines from investigators that select observations, and separate read-only investigation from write authority bounded by approval or a restricted action envelope.

\begin{table}[htbp]
\centering
\caption{Capability and evidence audit, part~2 of~3. This part continues the L1 systems from Table~\ref{tab:agent-systems} and begins the L2 read-only investigators.}
\label{tab:agent-systems-b}
\scriptsize
\begin{tabularx}{\textwidth}{P{1.45cm}P{2.45cm}P{3.30cm}P{2.30cm}Y}
\toprule
\textbf{Stage} & \textbf{System and venue} & \textbf{Capability and grounding} & \textbf{Evaluated authority} & \textbf{Strongest evidence and boundary} \\
\midrule
\multicolumn{5}{@{}l}{\textit{L1: copilots and fixed pipelines, continued from Table~\ref{tab:agent-systems}}} \\[1pt]
Diagnose & mABC~\cite{zhang2024mabc}; EMNLP Findings 2024 & Microservice RCA by a panel of role-specialized agents over supplied evidence. & Fixed collaboration and voting workflow; L1 advice. & E1: constructed benchmark; consensus damps but does not remove correlated error. \\
Localize & TrioXpert~\cite{sun2025trioxpert}; ASE 2025 & Detection, triage, and localization over metrics, logs, and traces. & Fixed modality-specific processing and aggregation pipeline; L1 analysis. & E1/E3: two datasets and reported deployment; no adaptive evidence query or mitigation loop. \\
Diagnose & SCELM~\cite{sun2025scelm}; FSE Comp. 2025 & Change assessment and triage from multimodal change data and retrieved knowledge. & Single automated retrieval and inference pipeline; L1 support. & E3: reported eleven-month operation; no iterative tool selection or action authority. \\
Localize & ThinkFL~\cite{zhang2026thinkfl}; TOSEM 2026 & Failure localization refined by reinforcement fine-tuning. & Self-refining learned localizer over supplied inputs; L1 analysis. & E1: benchmark datasets; refines model behavior, not an evidence-gathering policy. \\
\midrule
\multicolumn{5}{@{}l}{\textit{L2: read-only investigators, which select their own observations but change nothing}} \\[1pt]
Mitigate & LLexus~\cite{lascasas2024llexus}; SIGOPS OSR 2024 & End-to-end incident assistance from runbooks and telemetry. & Agent drafts and guides; human owns every change; L1/L2. & E3: deployed agent system; the human gate is a stated design constraint. \\
Diagnose & RCAgent~\cite{wang2024rcagent}; CIKM 2024 & Cloud RCA, solution, evidence, and owner using platform tools and domain knowledge. & Iterative tool use; L2 read-only investigation. & E3: integrated into an Alibaba Cloud workflow, not unrestricted remediation. \\
Diagnose & OpenRCA~\cite{xu2025openrca}; ICLR 2025 & RCA over 68+ GB of logs, metrics, and traces. & Retrieval/tool-assisted benchmark agent; L2 read-only. & E1: 335 failures; best reported solve rate 11.34\%. \\
Diagnose & Flow-of-Action~\cite{pei2025flow}; WWW Comp. 2025 & Microservice RCA grounded in telemetry, SOPs, and generated support artifacts. & SOP-guided multi-agent tool loop; L2 read-only. & E1 fault-injection testbed; improves over ReAct but does not validate production action. \\
\bottomrule
\end{tabularx}
\end{table}

\begin{table}[htbp]
\centering
\caption{Capability and evidence audit, part~3 of~3. This part completes L2, records the two write-capable systems, and lists an evaluation environment for comparison.}
\label{tab:agent-systems-c}
\scriptsize
\begin{tabularx}{\textwidth}{P{1.45cm}P{2.45cm}P{3.30cm}P{2.30cm}Y}
\toprule
\textbf{Stage} & \textbf{System and venue} & \textbf{Capability and grounding} & \textbf{Evaluated authority} & \textbf{Strongest evidence and boundary} \\
\midrule
\multicolumn{5}{@{}l}{\textit{L2: read-only investigators, continued from Table~\ref{tab:agent-systems-b}}} \\[1pt]
Diagnose & TAMO~\cite{zhang2025tamo}; IEEE TSC 2025 & Fine-grained RCA over multi-modality observation data. & Tool-assisted agent; L2 read-only. & E1/E2: cloud-native datasets; no mitigation authority. \\
Diagnose & FLASH~\cite{zhang2024flash}; MSR TR 2024 & Workflow automation for \emph{recurring} incident diagnosis. & Iteratively selects diagnostic tools; L2 read-only investigation. & E2/E3: 250 production incidents, owner evaluation, and reported deployment; tool outputs are stubbed during evaluation. \\
\midrule
\multicolumn{5}{@{}l}{\textit{L4: bounded closed loops with autonomous write authority inside a predefined envelope}} \\[1pt]
Mitigate & Nissist~\cite{an2024nissist}; ECAI 2024 & Mitigation grounded in troubleshooting guides and mitigation history. & Automatically executes supported actions and delegates unsupported ones; L4 bounded loop. & E3: user study with 20 on-call engineers and five incidents; no rollback analysis or controlled production outcome. \\
Mitigate & STRATUS~\cite{chen2025stratus}; NeurIPS 2025 & Detection through mitigation with Transactional Non-Regression under writer exclusivity, faithful undo, and bounded risk. & Autonomous multi-agent tool loop with rollback; L4 bounded closed loop. & E1: interactive benchmarks; the conditional guarantee depends on stated assumptions and is not measured in production. \\
\midrule
\multicolumn{5}{@{}l}{\textit{Listed for comparison: an evaluation environment rather than an operational system}} \\[1pt]
Detect to Mitigate & AIOpsLab~\cite{chen2025aiopslab}; MLSys 2025 & Detection through mitigation over live apps, faults, workloads, and telemetry. & Interface between an agent and the cloud; evaluation environment. & E1 interactive testbed; not itself an operational product. \\
\bottomrule
\end{tabularx}
\end{table}

Together, the rows show that current systems concentrate on evidence assembly. Fixed multi-model or multi-agent pipelines remain L1 when they neither choose new observations nor act on the service. At the write boundary, Nissist executes only actions supported by predefined tooling, whereas STRATUS constrains autonomous mitigation through transaction and rollback semantics. Neither carries E4 evidence. This distinction motivates the analysis of operational learning and evaluation that follows.

\subsection{Operational Learning}

Learning closes the loop when a verified incident outcome changes later observation, investigation, or execution. Using incident history as an input is useful, but is not itself learning, which explains the sparse learning column in Figure~\ref{fig:autonomy-matrix}.

DejaVu, iSQUAD, MonitorAssistant, FlowXpert, and X-lifecycle reuse historical failures, monitor records, or operational documents~\cite{li2022dejavu,ma2020isquad,yu2024monitorassistant,shi2025flowxpert,goel2024xlifecycle}. Their outputs remain a location, monitor, workflow, or diagnosis, so the history is an input to another stage rather than a verified update to future operations.

OpsMem, KRCA, OpsAgent, TopoEvo, and NetAgentMemory persist investigation state, procedures, or shared context~\cite{sun2026opsmem,jiang2026krca,luo2025opsagent,wang2026topoevo,ren2026netagentmemory}. Their updates still follow investigation trajectories rather than verified outcomes, and no study reports cheaper resolution on a later occurrence. Learning therefore remains read-only and unvalidated.

\subsection{Evidence Boundary of Current Systems}

The 22 audited systems make the authority boundary clear. Fourteen remain at L1, six reach L2 read-only investigation, none occupies supervised L3 operation, and two reach L4 bounded closed-loop operation. All write authority is concentrated in mitigation. Diagnosis accounts for 12 systems, and localization accounts for three. No audited language-model-enabled system is evaluated at learning. Twelve entries report E3 deployment or user-study evidence, but most of that evidence concerns advice or investigation. Nissist carries E3 evidence and automatically executes actions supported by its engine, while delegating unsupported actions to on-call engineers~\cite{an2024nissist}. STRATUS carries E1 evidence from interactive benchmarks and adds explicit rollback semantics~\cite{chen2025stratus}. Neither system carries E4 evidence. The only E4 result in the broader literature reviewed here is Narya's bounded preventive mitigation~\cite{levy2020narya}.

Evidence thins as authority rises because executed actions expose production to the error being measured. Narya's fifteen-month comparison and SCELM's eleven-month operational record approach outcome evidence by constraining the action or decision space~\cite{levy2020narya,sun2025scelm}; only Narya measures a direct operational outcome. The remaining gap is therefore an evaluation problem: benchmarks must connect authority to verified recovery and side effects, as discussed in Section~\ref{sec:evaluation}.

\section{Design Patterns for Agentic Service Operations}\label{sec:patterns}

The systems audited in Section~\ref{sec:systems} point to four recurring requirements for agentic service operations. They concern how evidence and system state are represented, how diagnosis proceeds under uncertainty, how agents interact with operational tools, and how changes are verified. The four parts below synthesize design requirements that recur across different architectures and connect them to the operational risks summarized in Table~\ref{tab:risks}.

\subsection{Evidence and State Representation}

Operational evidence should be converted into typed, time-bounded facts before it enters an agent's reasoning context. Each fact should identify the relevant entity, signal, time range, baseline, source query, and confidence, while retaining a link to the original record. LogPilot infers alert intent before retrieval, TrioXpert applies modality-specific preprocessing, and Xpert recommends a diagnostic query to scope retrieval~\cite{jiang2025logpilot,sun2025trioxpert,jiang2024xpert}. The grounding stage should also record the software version that produced the evidence. LogRobust shows that log formats change across releases~\cite{zhang2019logrobust}; without version information, a representational change may be mistaken for a change in system behavior.

The same evidence should be connected to a dynamic world model that relates business operations to their services, versions, workloads, pods, nodes, routes, owners, service-level objectives, and recent changes. Traces, Kubernetes state, and service-mesh traffic and policy provide complementary runtime views~\cite{sigelman2010dapper,burns2016borg,kosinska2023observability,song2024canalmesh}. MicroRCA constructs an attributed graph before ranking candidates, while Groot provides an event graph that engineers can inspect and amend~\cite{wu2020microrca,wang2021groot}. Temporal knowledge graphs and object-centric observability models place entities, relations, signals, and changes in a shared schema~\cite{zhang2024datasilos,pei2026umodel}. Missing observations should be represented explicitly so that well-instrumented components are not treated as more suspicious simply because they produce more evidence.

\subsection{Reasoning Under Uncertainty}

A diagnosis should maintain competing hypotheses and record the evidence supporting or contradicting each one, the symptoms it leaves unexplained, and the observation that would best distinguish it from alternatives. Causal graphs constrain admissible propagation paths~\cite{li2022circa,ikram2022rcd}, but comparative studies find no universal causal-RCA method and weak transfer from synthetic to real systems~\cite{pham2024causal}. Tool-calling agents often outperform retrieval-only assistants because they obtain additional evidence rather than because they reason better over identical inputs~\cite{roy2024llmagents}. The stopping condition should therefore depend on discriminating evidence, with calibrated confidence providing a basis for escalation when that evidence is insufficient~\cite{zhang2024lmpace}.

Runbooks, historical incidents, code, and configuration should be retrieved with their applicable services and versions, owner, validation time, and assumed preconditions. AutoTSG and Nissist show how troubleshooting guides can become executable or traversable knowledge~\cite{shetty2022autotsg,an2024nissist}. The key requirement is lifecycle management, since a procedure that resolved an earlier incident may become unsafe after an architectural or storage change. Agentic memory preserves intermediate reasoning states, while OpsMem separates current investigation records from distilled procedural knowledge~\cite{zhang2026agenticmemory,sun2026opsmem}. These stores should retain verified outcomes and expiry information, because an incorrect diagnosis can otherwise be retrieved as support for a later incident.

\subsection{Governed Interaction}

Every operational tool should declare its input schema, semantic description, execution identity, timeout, cost, and side-effect class. A broker between the agent and the platform can validate parameters, remove secrets, record provenance, enforce budgets, and apply policy before execution~\cite{shetty2024autonomousclouds,chen2025aiopslab}. Read tools should be the default. Test tools may run health checks, replay traffic, or conduct pre-approved experiments, following the discipline of bounded chaos experiments~\cite{basiri2016chaos}. Write tools require stronger controls for target selection, approval, canary scope, postconditions, and rollback. Policy engines provide a reviewable and versioned enforcement point~\cite{opa2026}. Telemetry can also contain instructions that manipulate tool selection, so it must be treated as untrusted input and isolated from operational instructions~\cite{greshake2023injection,he2025agentsecurity}.

Multiple agents are justified when their data, tools, permissions, or validation responsibilities differ. Flow-of-Action separates observation from stopping decisions, TrioXpert assigns expertise by telemetry modality, and STRATUS separates read-only agents from a repair agent with rollback support~\cite{pei2025flow,sun2025trioxpert,chen2025stratus}. COLA applies the same principle by using correlation for easy cases and an LLM only for ambiguous pairs~\cite{kuang2024cola}. Conversational frameworks, encoded procedures, and explicit communication topologies make composition easier to implement~\cite{wu2024autogen,hong2024metagpt,li2023camel,guo2024multiagentsurvey}. Yet agent count is not evidence of better diagnosis: agentless pipelines can match or exceed agent scaffolds at lower cost~\cite{xia2025agentless}. A decomposition should therefore create independent evidence or validation paths, rather than several agents that share the same context and failure mode.

\subsection{Outcome Assurance}

Operational change should be preceded by a precondition query, dependency and impact analysis, a preview or dry run when available, risk-based approval, limited-scope execution, and explicit postconditions over service-level objectives and health signals. Rollback should be triggered when those postconditions fail. A successful API response does not show that the service recovered. Narya provides the clearest reference because it combines bounded preventive actions with automatically measured production outcomes~\cite{levy2020narya}. STRATUS expresses a similar target through Transactional Non-Regression under writer exclusivity, faithful undo, and bounded risk~\cite{chen2025stratus}. Authority should therefore be granted according to verified recovery and side effects, rather than the number of agents or the fluency of their explanations. Table~\ref{tab:risks} summarizes the risks addressed by these four requirements and the guardrails used to contain them.

\begin{table}[htbp]
\centering
\caption{Failure modes addressed by the four design requirements. Each row links an operational risk to the signal that exposes it and the guardrail that contains it.}
\label{tab:risks}
\small
\begin{tabularx}{\textwidth}{P{4.0cm}P{3.8cm}Y}
\toprule
\textbf{Failure mode and manifestation} & \textbf{Detection signal} & \textbf{Guardrail} \\
\midrule
Unsupported narrative: plausible cause without traceable telemetry. & Claims lack source IDs or conflict with observed facts. & Evidence ledger; separate fact, inference, and recommendation. \\
Stale knowledge: obsolete command, resource, or dependency assumption. & Version or scope mismatch; old validation timestamp. & Versioned retrieval with applicability and freshness checks. \\
Context overload: rare but decisive evidence is ignored. & Repeated broad queries, token growth, or unstable answers. & Intent scoping, clustering, and summaries linked to raw records. \\
Wrong/excessive tool use: costly scans, irrelevant probes, repeated loops. & Budget exhaustion or low information gain. & Typed catalog, per-tool budgets, and loop/stop policies. \\
Unsafe write: scaling, routing, or rollback increases impact. & Failed precondition or SLO degradation after canary. & Least privilege, approval, canary, postcondition, and rollback. \\
Prompt/data injection: telemetry text manipulates tool selection. & Untrusted text requests secrets, instructions, or commands. & Treat telemetry as data; isolate instructions and filter output. \\
Correlated agent error: agents share the same false assumption. & Agreement without independent evidence. & Heterogeneous checks, counterevidence role, deterministic validator. \\
Automation bias: a fluent but uncertain diagnosis is accepted. & Low review time or weak evidence inspection. & Calibrated uncertainty, forced evidence view, and easy escalation. \\
\bottomrule
\end{tabularx}
\end{table}

\section{Evidence-Oriented Evaluation of Agentic Service Operations}\label{sec:evaluation}

The audit in Section~\ref{sec:systems} showed that evidence weakens as evaluated authority rises. This section connects resources, metrics, settings, and validity threats to the authority exercised, rather than treating benchmark scores as evidence of autonomy.

\subsection{Datasets and Interactive Environments}

Table~\ref{tab:benchmarks} states the strongest claim and main limitation of each resource. Static datasets, e.g.,\ Loghub, RCAEval, and OpenRCA, support reproducible diagnosis but not stateful investigation or recovery~\cite{zhu2019logparser,pham2025rcaeval,xu2025openrca}. OpenRCA~2.0 adds stepwise causal labels~\cite{fang2026openrca2}, while DeathStarBench provides configurable workloads whose ground truth still depends on experiment design~\cite{gan2019deathstar}.

Interactive environments expose tools, changing state, and outcomes. AIOpsLab covers detection through mitigation~\cite{chen2025aiopslab}; ORCA-bench demonstrates that available source-code evidence changes every metric~\cite{gong2026orcabench}; and ITBench evaluates live SRE, FinOps, and compliance tasks~\cite{jha2025itbench}. AIOps2025/RCA100 adds expert-scored reasoning traces~\cite{cai2026agenticopseval}. Broad detection suites improve domain coverage~\cite{paparrizos2022tsbuad}, while fault-injection frameworks improve control but cannot reproduce the distribution of organic incidents~\cite{chaosmesh2026}.

\begin{table}[htbp]
\centering
\caption{Evaluation resources and the claims they can support.}
\label{tab:benchmarks}
\small
\begin{tabularx}{\textwidth}{P{3.0cm}P{2.9cm}P{3.05cm}Y}
\toprule
\textbf{Resource and content} & \textbf{Interaction mode} & \textbf{Strongest supported claim} & \textbf{Important limitation} \\
\midrule
Loghub~\cite{zhu2019logparser}: logs from multiple systems. & Single-modal and static; no tool or action loop. & Log parsing and detection generalization. & No service topology or incident action. \\
DeathStarBench~\cite{gan2019deathstar}: microservice applications and workloads. & Live, configurable experiment harness. & System behavior under controlled workloads and injected faults. & Ground truth depends on experiment design. \\
RCAEval~\cite{pham2025rcaeval}: 735 cases, three systems, fifteen baselines. & Multimodal and static; no tool or action loop. & Comparable coarse- and fine-grained RCA ranking. & Static cases do not test investigation policy. \\
OpenRCA~\cite{xu2025openrca}: 335 failures and 68+ GB telemetry. & Multimodal replay with read/retrieve interaction. & Long-context, evidence-grounded agentic RCA. & No mitigation outcome. \\
OpenRCA 2.0~\cite{fang2026openrca2}: 500 cases with stepwise causal labels. & Multimodal replay; intermediate causal steps are scored. & Whether the causal chain, not only the named service, is correct. & Chains are derived from injected faults; no mitigation outcome. \\
ORCA-bench~\cite{gong2026orcabench}: 1{,}079 graded on-call tasks on a live system. & Live, instrumented environment with code and telemetry access. & Readiness of coding agents for on-call diagnosis. & Single testbed; difficulty grading is benchmark-defined. \\
AIOpsLab~\cite{chen2025aiopslab}: live apps, workloads, and fault library. & Live multimodal state, tools, and action checks. & Interactive incident-lifecycle performance. & Injected faults may be cleaner than organic failures. \\
AIOps2025\slash RCA100~\cite{cai2026agenticopseval}: 500+ expert-labeled cases. & Multimodal and static; scored reasoning trace; no action. & Localization, fault identification, and evidence reasoning. & Competition labels do not prove safe action. \\
ITBench~\cite{jha2025itbench}: SRE, FinOps, and compliance scenarios. & Live environments with agent tooling and scored outcomes. & Cross-domain agent competence on real IT tasks. & Reported resolution rates are low; scenario coverage is uneven. \\
TSB-UAD~\cite{paparrizos2022tsbuad}: curated univariate anomaly suite. & Single-modal and static; detection only. & Detection generalization across domains and anomaly types. & No service context, topology, or action. \\
\bottomrule
\end{tabularx}
\end{table}

\subsection{Four Metric Families}

Evaluation should report four metric families: outcome quality, e.g.,\ detection, ranking, and recovery; evidence quality, including support, contradictions, and causal completeness; trajectory quality, including tool validity, cost, latency, and stopping; and safety and human factors, including unauthorized actions, rollback, takeover, and usefulness. Together they connect an answer to the evidence and actions that produced it.

These families must be stage-specific. Detection requires event- and point-level precision and recall; localization requires rank-sensitive measures; diagnosis requires both causal correctness and evidential support; and mitigation requires recovery, rollback, and collateral-effect measures. Aggregate success rates should therefore be accompanied by coverage, abstention, resource consumption, and human-intervention frequency. Otherwise, a system can improve apparent accuracy by attempting only easy incidents, issuing excessive queries, or transferring unresolved work to operators. Reporting both attempted and eligible cases makes these trade-offs visible.

Protocols can nevertheless create misleading progress. Point adjustment, benchmark triviality, and mislabelling have made weak time-series detectors appear competitive~\cite{kim2022rigorous,wu2023currenttime}, and careful baselines have challenged complex forecasting and log-parsing models~\cite{zeng2023transformers,jiang2024loghub2}. Final-answer scoring creates an analogous problem for agents that guess before searching. In OpenRCA~2.0, $76.0\%$ correct-service accuracy falls to $61.5\%$ for a verified path and $20.7\%$ for the complete root-cause set~\cite{fang2026openrca2}; ORCA-bench reports hallucinated causes for up to $40\%$ of cases~\cite{gong2026orcabench}. Diagnosis and remediation therefore require execution checks and evidence attribution, not text similarity alone.

\subsection{An Evaluation Ladder}

\begin{figure}[tbp]
\centering
\includegraphics[width=\linewidth]{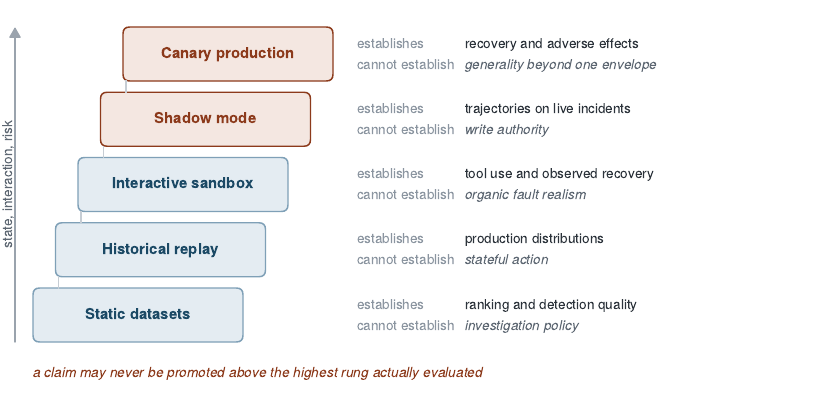}
\caption{The evaluation ladder, with what each rung can and cannot establish. Higher rungs do not replace lower ones; they add state, interaction, and risk. The right-hand column is the operative one: a claim may never be promoted above the highest rung actually evaluated.}
\label{fig:evaluation-ladder}
\Description{Five ascending steps run from static datasets through historical replay, interactive sandbox and shadow mode to canary production. Each step is annotated with what it establishes and what it cannot establish, and a vertical axis marks increasing state, interaction and risk.}
\end{figure}

Figure~\ref{fig:evaluation-ladder} orders settings by state, interaction, and risk. Static data support offline analysis; replay adds production distributions but retains hindsight; sandboxes test tools and recovery in controlled state; shadow mode observes live incidents without writes; and canary production adds narrow, reversible action. Claims must remain within the rung actually evaluated.

\subsection{Threats to Validity}

Four threats recur. Injected faults differ from gradual, concurrent, policy-induced, and external incidents; agents may recognize benchmark conventions; production datasets overrepresent detected and documented cases; and human judgments depend on expertise and visible evidence. Model and tool drift compounds these threats. Reproducible studies should record versions, prompts, schemas, knowledge snapshots, permissions, seeds, trajectories, and exercised authority. The central gap is evidence connecting investigation and authority to recovery and side effects.

These threats interact rather than occur independently. A benchmark-contaminated model can exploit an injected-fault pattern, produce a plausible explanation, and still fail when a provider dependency or policy interaction causes the same symptoms. Likewise, an expert may accept a recommendation because its cited records look familiar even when the collection policy omitted contrary evidence. Evaluations should therefore separate dataset construction, evidence collection, reasoning, and outcome adjudication; report disagreements and excluded incidents; and replay representative failures after model, tool, or knowledge updates. Longitudinal replication is particularly important for agentic systems because the executable interface can change even when the model name and benchmark remain fixed.

\section{Open Problems and Research Agenda}\label{sec:agenda}

The weakly graded regions of Figure~\ref{fig:autonomy-matrix} expose four connected problems: grounding a testable service model, reasoning under incomplete evidence, acting within a controlled envelope, and demonstrating operational improvement.

\subsection{Grounding the Operational World}

Collecting everything is neither feasible nor desirable; observability correlates heterogeneous partial views~\cite{kosinska2023observability}. Agents should request extra traces, profiles, or logs only when they distinguish hypotheses. Alert-scoped retrieval and query recommendation are early forms of this policy~\cite{jiang2025logpilot,jiang2024xpert}, which must price telemetry, privacy, and instrumentation risk. Privacy budgets~\cite{kong2020privacy}, restricted-access diagnosis~\cite{zhai2025nrcac}, and uncertainty-triggered labels~\cite{sun2026uncertainty} provide useful components. The resulting problem is adaptive sensing.

Such sensing also needs an explicit observation ledger. Each request should record the hypothesis it tests, expected information gain, time range, source version, cost, and whether the returned evidence was complete. This record separates a justified absence from missing telemetry and allows later evaluators to distinguish an unsuccessful investigation policy from an unobservable incident. It also exposes repeated queries and uncontrolled sampling increases that a final-answer metric would hide.

Those observations must ground a testable world model. Causal RCA degrades when inferred or supplied structure is incomplete~\cite{li2022circa,ikram2022rcd,zheng2024mulan,pham2024causal,xie2024latentscope}. An agent could instead select a safe probe, compare counterfactual predictions, and update the model; bounded chaos, digital twins, replay, shadowing, and canaries offer candidate mechanisms~\cite{basiri2016chaos}. Portability further requires shared entities and relations~\cite{deng2025asc,pei2026umodel}. General agent protocols support discovery and delegation~\cite{mcp2025spec,a2a2025spec}, but operational tool contracts must also expose side effects, reversibility, blast radius, and postconditions. Reproducible schemas and traces would enable cross-platform comparison.

\subsection{Reasoning Beyond Known Cases}

Benchmarks commonly assume one injected fault and a known candidate set~\cite{pham2025rcaeval,chen2025aiopslab}. Production also contains interacting faults, metastability, external dependencies, and causes outside the service graph, helping explain low solve rates on long-context data~\cite{xu2025openrca}. Generative-AI services add quota, model, prompt, and serving-stack failures~\cite{yan2025genaiincidents}. Agents therefore need open-set recognition, insufficient-evidence reports, competing hypotheses, and useful abstention rather than forced labels~\cite{chang2026causaldx}.

Open-set evaluation should include incidents whose true cause is absent from the candidate graph, as well as cases in which several causes remain observationally equivalent. A calibrated escalation is then a correct operational outcome rather than a failed prediction. Benchmarks should score whether the agent identifies the missing evidence and selects a safe next probe, not merely whether its final explanation resembles a reference answer.

Runbooks and incident histories encode expertise but also obsolete architecture and mistakes~\cite{shetty2022autotsg,an2024nissist,zhang2019logrobust}. Memory systems separate current state, reusable experience, and shared context~\cite{sun2026opsmem,zhang2026agenticmemory,ren2026netagentmemory}, yet none reports reliable expiry and outcome gating. Temporal retrieval, applicability checks, ownership, and verified post-incident outcomes are necessary to prevent an incorrect answer from becoming evidence for the next incident.

\subsection{Constraining Operational Action}

The frontier is a library of reversible, high-level actions with machine-checkable contracts, not broader shell access. Policy-as-code, least privilege, risk budgets, canaries, and rollback must be evaluated. Constrained response selection illustrates a narrow envelope~\cite{zhang2026honeypoint}; Narya couples one with production outcomes~\cite{levy2020narya}; and STRATUS broadens it through rollback-based Transactional Non-Regression under explicit assumptions~\cite{chen2025stratus}. Higher authority is thus more likely to comprise many narrow envelopes than one universal operator.

Responsibility remains organizational. Overtrust, disuse, and loss of routine practice are established automation risks~\cite{parasuraman1997use,bainbridge1983ironies}. Studies should measure toil, bias, expertise, and accountability; Nissist's user study remains unusual~\cite{an2024nissist}. Timelines, dependency graphs, hypothesis boards, and change diffs can expose uncertainty better than chat alone. Human corrections should update structured evidence and evaluation cases, while ranked feedback can make operator judgment reusable~\cite{wang2023rcahrl}.

\subsection{Establishing Evidence for Progress}

Realistic final-answer accuracy remains low~\cite{xu2025openrca}, while interactive environments, expert-scored traces, stepwise causal labels, and stage-attributed trajectories expose how answers were reached~\cite{chen2025aiopslab,cai2026agenticopseval,fang2026openrca2,wang2026star}. Given known protocol failures in anomaly detection~\cite{kim2022rigorous}, benchmarks should score evidence selection, counterevidence, stopping, unauthorized attempts, recovery, and side effects. Each operational result should form an assurance case containing the objective, facts, causal argument, alternatives, policy checks, approved action, and verified outcome.

Publishing the assurance case enables failure analysis that a scalar score cannot provide. Researchers can identify whether an error arose from missing telemetry, an invalid inference, a stale runbook, an unsafe tool choice, or an incorrect recovery check. It also permits asymmetric scoring: unsupported write attempts should carry a larger penalty than redundant read calls, while calibrated abstention should be rewarded when decisive evidence is unavailable. Shared schemas for these records would make results comparable across datasets, simulators, and production studies without assuming that all platforms expose identical tools.

\section{A Controlled Adoption Path}\label{sec:practice}

Adoption should begin with verifiable evidence and narrow authority. A service catalog must connect operations to services, versions, deployments, workloads, routes, owners, and SLOs, while telemetry retains provenance. RCACopilot deployed information collection before introducing an LLM~\cite{chen2024rcacopilot}, and Groot relied on a maintainable event graph~\cite{wang2021groot}. The first action target should be frequent, reversible, and automatically testable; Narya follows this rule with a bounded mitigation set and nine production experiments~\cite{levy2020narya}.

Historical replay should compare rules, RCA models, and human timelines using evidence quality, cost, abstention, and time, not only labels~\cite{chen2024rcacopilot,zhang2024iclrca,ahmed2023recommending,kim2022rigorous}. Read-only shadow mode can then investigate live alerts while humans retain write authority. Existing deployments span alert diagnosis, change assessment, triage, and platform-integrated investigation~\cite{jiang2025logpilot,sun2025scelm,wang2024comet,kuang2024cola,wang2024rcagent}.

Supervised actions require contracts for preconditions, previews, approval, postconditions, and rollback~\cite{an2024nissist,chen2025stratus}. Credentials and general shells should remain outside model context because the agent-platform interface enforces this boundary~\cite{shetty2024autonomousclouds}. Only repeated shadow and canary evidence should justify a bounded loop; promotion follows exercised authority, not model capability~\cite{parasuraman2000levels}. Kill switches and human routing remain necessary for novel cases~\cite{bainbridge1983ironies}.

Verified incidents should become regression tests, benchmark cases, governed monitor or runbook updates~\cite{yu2024monitorassistant}, and expiring knowledge records. Before raising authority, a review should answer six questions: which entities and time ranges ground each claim; which tools the agent can call; which calls mutate state; how uncertainty and abstention are represented; what verifies recovery and triggers rollback; and which evaluation rung supports the requested authority. These answers should be machine-readable where possible, since permissions, tool contracts, and service ownership change more quickly than model prompts. Promotion should be revoked when the evidence, interface, or operating envelope changes materially.

Promotion is therefore a reversible operational decision, not a permanent property of a model. Teams should monitor abstention, human overrides, rollback frequency, investigation cost, and incident classes outside the approved envelope. A rise in unsupported actions or a change in telemetry coverage should return the system to shadow mode until replay and canary evidence are re-established. The same evidence ledger used during incidents can support this review by showing which claims depended on changed schemas, tools, or knowledge. This closes the adoption loop without allowing unverified incident summaries to become future operating policy.

\section{Conclusion}\label{sec:conclusion}

This survey organizes the transition from AIOps pipelines to agentic service operations around service substrates, the incident lifecycle, operational authority, and evidence. Its taxonomy, four-plane architecture, evidence grades, and evaluation ladder place analytical methods, copilots, and tool-using agents in one operational frame, clarifying which claims their permitted actions and evaluations support.

Multimodal analysis, causal diagnosis, retrieval, and tool-assisted investigation have advanced, while interactive environments and stepwise benchmarks increasingly expose investigation behavior~\cite{chen2025aiopslab,fang2026openrca2}. Bounded mitigation also shows how production outcomes can be measured~\cite{levy2020narya}. Progress now depends on connecting typed service state and evidence-grounded reasoning to governed actions, verification, rollback, and human oversight. Authority should increase only when evaluation demonstrates recovery and acceptable side effects at that same level.

\bibliographystyle{ACM-Reference-Format}
\bibliography{csur}

\end{document}